\documentclass[longbibliography, notitlepage, prd, superscriptaddress,amssymb,nofootinbib,amsmath,showkeys]{revtex4-1}
\pdfoutput=1

\usepackage{graphicx}
\usepackage{psfrag}
\usepackage{enumerate}

\newcommand{\be}{\begin{equation}}
\newcommand{\ee}{\end{equation}}
\newcommand{\beq}{\begin{eqnarray}}
\newcommand{\eeq}{\end{eqnarray}}

\def\[{\left [}
\def\]{\right ]}
\def\({\left (}
\def\){\right )}

\def\r2{\sqrt{2}}

\def\n{{\bf \hat{n}}}

\newcommand{\bbibitem}[1]{\bibitem{#1}\marginpar{#1}}

\newcommand{\figref}[1]{Fig. \ref{#1}}
\newcommand{\secref}[1]{Sec. \ref{#1}}

\def\Label#1{\label{#1}%
  \smash{\hbox to0pt{\raise1ex\hbox{\tiny[#1]}\hss}}}
\def\noLabels{\let\Label=\label}
\def\nobbibitem{\let\bbibitem=\bibitem}

\begin{document}

\noLabels
\nobbibitem

\DeclareGraphicsExtensions{.pdf,.png,.gif,.jpg,.eps}

\title{Polarizing Bubble Collisions}

\author{Bart{\l}omiej Czech}
\email{czech@phas.ubc.ca}
\affiliation{Department of Physics and Astronomy, University of British Columbia, Vancouver, BC V6T 1Z1, Canada}

\author{Matthew Kleban}
\email{mk161@nyu.edu}
\affiliation{CCPP, Department of Physics, New York University, New York, NY 10003, USA}

\author{Klaus Larjo}
\email{larjo@phas.ubc.ca}
\affiliation{Department of Physics and Astronomy, University of British Columbia, Vancouver, BC V6T 1Z1, Canada}

\author{Thomas S. Levi}
\email{tslevi@phas.ubc.ca}
\affiliation{Department of Physics and Astronomy, University of British Columbia, Vancouver, BC V6T 1Z1, Canada}

\author{Kris Sigurdson}
\email{krs@phas.ubc.ca}
\affiliation{Department of Physics and Astronomy, University of British Columbia, Vancouver, BC V6T 1Z1, Canada}


\begin{abstract}
\noindent
We predict the polarization of cosmic microwave background (CMB) photons that results from a cosmic bubble collision.  The polarization is purely $E$-mode, symmetric around the axis pointing towards the collision bubble, and has several salient features in its  radial dependence that can help distinguish it from a more conventional explanation for unusually cold or hot features in the CMB sky.  The anomalous ``cold spot'' detected by the Wilkinson Microwave Anisotropy Probe (WMAP) satellite is a candidate for a feature produced by such a collision, and the Planck satellite and other proposed surveys will measure the polarization on it in the near future.  The detection of such a collision would provide compelling evidence for the string theory landscape.

\end{abstract}






\maketitle

\begin{table}[t]
\begin{tabular}{|l|l|}
 \hline
 \multicolumn{2}{|c|}{\bf  Symbols and Definitions} \\
  \hline
  $T_{0}$ & Average photon temperature today ($\sim 2.726$ K)\\
  $T_{dc}(x,y,w)$ & Photon temperature at decoupling ($t=t_{dc}$)\\
  $D_{dc}$ & Comoving distance from Earth to the decoupling/recombination surface \\
  $D_{e}$ & Comoving distance to a scattering electron \\
  $\alpha$ & The ratio $D_{e}/D_{dc}$ \\
  $x_{c}$ & Comoving location of the collision lightcone at decoupling\\
 $\tilde \lambda$ & Linear slope of the temperature perturbation at decoupling \\
  $\lambda$ & Slope of the ``effective" perturbation including Sachs-Wolfe (see text) \\ 
  $\theta_{c}$ & Angular radius of the affected disk on the CMB temperature map \\
  $\mu$ & $\cos\theta$ \\
  \hline \end{tabular}
\end{table}

\section{Introduction}

Recently, considerable attention has been focused on the question of the observability of cosmic bubble collisions.  These collisions occur in models in which our entire Hubble patch---in fact, the entire region of the Universe in our vicinity describable by an approximately homogeneous and isotropic metric---exists inside a bubble created by a first-order phase transition from an eternally inflating false vacuum, and our bubble is struck by another that nucleated (from our parent false vacuum) nearby.  They are of interest in part because their existence is predicted by the string theory landscape \cite{discretum, lennylandscape} and is intimately related to its solution of the cosmological constant (CC) problem.

In the landscape the CC problem is solved by anthropic selection; that is, we find ourselves in a region of small vacuum energy because such phases exist in some places in the landscape, and it is only inside bubbles of these phases that structures like stars and galaxies can form \cite{Weinberg:1987dv}.  One expects such small CC bubbles to be created by nucleation from a rapidly inflating false vacuum.  If so, and if the instanton mediating the decay has the symmetries of the Coleman-de Luccia solution \cite{cdl}, the inside of the bubble is homogeneous and isotropic with negative spatial curvature (it contains inside it a complete, spatially infinite open Friedman-Robertson-Walker Universe).  Viewed from the outside the bubble expands at nearly the speed of light, but remains immersed in a bath of even more rapidly inflating false vacuum.  This phase is necessarily metastable and will occasionally decay to form other bubbles.  When another bubble---either of the same type or another---nucleates within one false-vacuum Hubble length of the wall of our bubble it collides with it, leaving an interesting and highly characteristic imprint on various large-scale cosmological observables \cite{wwc2,ktflow}.

Modulo the (substantial) theoretical uncertainties inherent in the above paradigm, the existence of such collisions is a prediction of string theory.  Hence the possibility of observing such an event presents an opportunity to greatly enhance our understanding of both cosmology on the largest possible length scales and of fundamental microphysics---and as such, it should be pursued vigorously.   The results are of interest outside of the context of string cosmology as well because the existence of such collisions is generic in any theory with slow-first order phase transitions coupled to gravity.  In addition, the techniques developed to analyze these collision spacetimes are of utility for the study of anisotropic and inhomogeneous cosmologies more generally.

\subsection{Probabilities}

In general, the expected number of potentially observable bubble collisions with bubbles of type $j$ was estimated in \cite{bubmeas}; it is
\be
\langle N_{c} \rangle \sim \gamma_j {V_{f} \over V_{i}}  \sqrt{\Omega_{k}},
\ee
where $\gamma_j$ is the decay rate of the false vacuum to vacua of type $j$ (in units of the false vacuum Hubble rate), $V_{f}$ is the parent false vacuum energy density, $V_{i}$ is the energy density during slow-roll inflation in our bubble, and $\Omega_{k}$ is the magnitude of the negative spatial curvature in our bubble today.  This is the number of collisions with future lightcones that bisect the part of the last scattering surface visible in the Cosmic Microwave Background (CMB) sky today (collisions whose future lightcones have not yet affected the CMB are unobservable, and those that encompass our entire sky and more are probably hard to detect).  This number could be significantly greater than one without fine-tuning (although it appears unlikely to be of order one).

Among the theoretical uncertainties is our ignorance of the decay rates of our parent false vacuum.  If all the rates are extremely slow, the spacetime region in our bubble that we can observe today  (despite its age of $\sim13.7$\,Gyr) may not yet have been affected by a collision with another bubble.  Another caveat is the amount of slow-roll inflation in our bubble.  After $N$ e-folds of slow-roll inflation, the spatial curvature is $\Omega_{k}\sim e^{2(N^{*}-N)}$, where $N^{*}$ is $\sim\!60$ (up to logarithmic dependence on factors like the reheating temperature (see {\it e.g.} \cite{ll}), so increasing the number of e-folds of inflation by $\Delta N$ decreases $\sqrt{\Omega_{k}}$ by a factor of $e^{-\Delta N}$.  Inflation also reduces the magnitude of the effects even for bubbles that bisect the CMB sky (by decreasing the brightness of the perturbed disk they create, for example). Hence large amounts of inflation both increase the distance between collisions by expanding our bubble and reduce the magnitude of their effects by inflating away their remnants.  Therefore, like any pre-inflationary relic, bubble collisions become exponentially harder to observe as the number of e-folds of inflation increases.

However, it appears that large numbers of e-folds of inflation require fine-tuning to attain in string theory, and at the same time that there is an anthropic requirement that there be sufficient inflationary expansion to solve the flatness problem \cite{Freivogel:2005vv}.  One then expects that the number of e-folds of inflation was not too much larger than the required minimum.  The theoretical uncertainties are too great to say anything more definite, but it does not appear that a large degree of fine-tuning is required for the effects of these collisions to be observable.

\subsection{Previous Work}
The cosmological implications of bubble formation and collision were studied recently in \cite{Freivogel:2005vv, ggv, ben, Aguirre:2007an, wwc, Aguirre:2007wm, wwc2, Aguirre:2008wy, bubmeas, ktflow,Easther:2009ft, Giblin:2010bd}.   Specifically, \cite{wwc2} derived the approximate CMB temperature anisotropy due to a collision---the effect is confined to a disk that is either hot or cold relative to the average, with an intensity that decreases linearly with cosine of the angular radius from the center of the disk, reaching zero at its edge.  That result will be our starting point here, but rather than reproducing the analysis of \cite{wwc2} we will begin with the collision temperature perturbation at decoupling and study its effects on CMB polarization.  The effect of the collision on the peculiar velocities of large scale structures was analyzed in \cite{ktflow}.  Previous work on CMB polarization and the WMAP cold spot appeared in \cite{polspot}.

\subsection{Summary of Results}
We compute the contribution to CMB polarization from a cosmic bubble collision, treating the collision's effects using a technique based on \cite{wwc2}.  We find that the polarization is purely $E$-mode (as expected for a scalar perturbation), and is radial (azimuthal) for collisions that produce cold (hot) disks in the temperature map.  If the polarization resulted entirely from Thomson scattering off electrons at a single value of the redshift $z$, the effects would be confined to an annulus containing the edge of the affected disk in the temperature map.  With a realistic model for scattering, the polarization is largest near the edge of the temperature disk, and drops to zero at its center and as one moves radially out from the edge.  

Our result is accurate for collision disks larger than a few degrees, but cannot be relied on quantitatively for features at sub-degree scales (see \secref{sec-analytic} for details).  If the WMAP cold spot  \cite{coldspot1,coldspot2,coldspot3,coldspot4,coldspot5} is the result of a bubble collision, the corresponding polarization pattern is strong enough to be detected by the Planck satellite.  We do not analyze the question of with what confidence one could rule out a random Gaussian origin were the measured polarization to be consistent with our prediction.  However,  we do emphasize that the temperature and polarization pattern we predict has a striking planar symmetry in three dimensions (inherited from the physics of the collision) which may be difficult to mimic with random fields.

\section{Bubble Collision Basics}
\label{setup}

\begin{figure}[t]
\begin{center}
\includegraphics[width=0.8\textwidth]{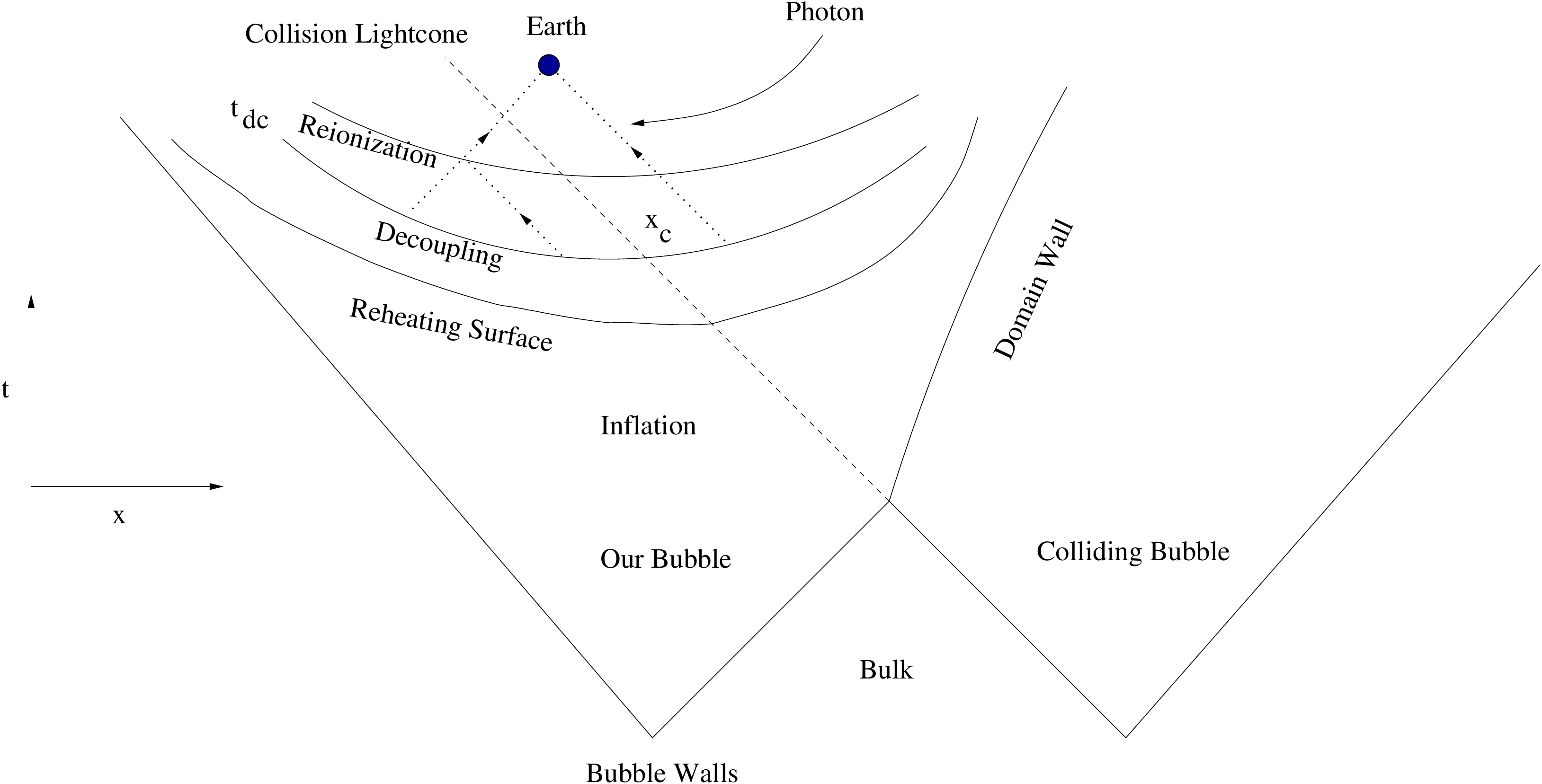}
\end{center}
\caption{\label{fig-setup} Sketch of the causal structure of the spacetime in the bubble collision scenario. The region to the right of the radiation shock is within the forward lightcone of the collision and affected by it.  The figure shows the trajectories of two photons that last scattered at decoupling, and one that last scattered at reionization.  The figure is not to scale.}
\end{figure}

We focus on a collision scenario where the bubble we inhabit formed by a Coleman-de Luccia transition from an eternally inflating parent vacuum, was struck by a bubble of a third type, and underwent a period of slow-roll inflation soon after its formation. This is schematically shown in \figref{fig-setup}. We assume that the pressures on the domain wall formed between the two colliding bubbles are such that the wall accelerates away from our bubble. This assumption is valid for collisions between two de Sitter (dS) bubbles if ours has a smaller vacuum energy, and  in collisions with anti-de Sitter bubbles when certain conditions involving the tension of the wall and the vacuum energies are satisfied \cite{ben,wwc,Aguirre:2007wm,Cvetic:1996vr}.  While more general scenarios are possible, these assumptions can lead to a cosmology consistent with current observations.  Finally, we assume that the collision lightcone divides the part of the decoupling surface that we can see in the CMB sky.

We wish to investigate the polarization of CMB photons induced by the collision.  A complete treatment would include the non-linear evolution of the cosmology from the highly anisotropic and inhomogeneous initial conditions set up by the collision. However, since observations constrain the magnitude of the collision signal to be small  in the region of the universe we can observe today, perturbation theory is sufficient to characterize its effects on most (if not all) signals.  The precise formulation of cosmological perturbation theory in this context and a detailed analysis of the evolution of cosmological perturbations in such a bubble collision Universe will be discussed in future work \cite{aniso}.

In this paper we instead make several approximations so that an analytic analysis that captures the key physics is possible.  We will treat the effects of the collision starting from the time $t=t_{dc}$ of decoupling between photons and atoms (also known as the time of recombination).  We assume that inflation lasted long enough so that the negative spatial curvature can be ignored, so that to leading order the background metric is approximately
\be
ds^{2} = -dt^{2} + a^{2}(t)  d{ \bf x}^{2} .
\ee
Here ${\bf x} = (x,y,w)$ is a comoving coordinate, the $x$-axis points towards the collision, and the lightcone of the collision event is a plane extended in the $y$ and $w$ directions, and null in the $x$ and $t$ directions (\figref{fig-setup}).  Without loss of generality the coordinate position of the Earth today is taken to be ${\bf x}=(0,0,0)$ and $t=t_{0}$.  

To model the effects of the collision, we take the perturbation to the temperature of the Universe due to the collision to be
\be
\label{3dtemp}
T_{dc}({\bf x}) = (1+z_{dc})T_{0}\left[ 1 +  \tilde \lambda (x-x_{c}) \Theta(x-x_{c}) \right]
\ee
at $t=t_{dc}$. Here $\Theta(x)$ is a step function, $x=x_{c}$ is the comoving coordinate of the collision lightcone at decoupling (see \figref{fig-setup}), and $\tilde \lambda$ is a parameter that controls the intensity of the affected disk in the CMB sky. The temperature perturbation is assumed to be adiabatic with no isocurvature contribution. The temperature in the part of the Universe unaffected by the collision is $T_{dc}=T_{0} (1+z_{dc})$, where $T_{0} \simeq 2.726$ K and $z_{dc}\simeq 1089$.  As we demonstrate below, this temperature profile reproduces the results of \cite{wwc2} for the CMB temperature perturbation due to a bubble collision.

So long as the perturbations are small, they may be treated linearly, and (as in \cite{wwc2}) we will assume that the random fluctuations produced during slow-roll inflation remain Gaussian and uncorrelated with the collision perturbation at lowest order.  As such, we will not discuss the usual random contribution to anisotropies but they will also be present with their usual amplitude.

Note that for large $x-x_{c}$ the effects of the collision become large and cannot be treated perturbatively.  This is due to the fact that near the domain wall separating the bubbles, the spacetime is strongly affected and cannot be approximated by a perturbed FRW metric.  However, consistency with the observed isotropy of the CMB requires that the perturbation be small in the part of the decoupling surface that falls within our past lightcone today, and as long as that remains true our approximation is valid.

The ansatz adopted here differs somewhat from that of \cite{wwc2, ktflow}.  Roughly speaking, \cite{wwc2, ktflow} treated reheating and decoupling as surfaces of constant temperature but non-constant FRW time, while here we treat them as surfaces of constant time but non-constant temperature (and gravitational potential).  To linear order in the perturbation these approaches are very similar (if not simply gauge-equivalent).  The end result of all these analyses is a perturbation with a spatial dependence that is dictated by the symmetries of the bubble collision and two parameters (the distance from the Earth's comoving location $x=0$ to the collision lightcone $x=x_c$, and the slope of the effect inside the lightcone $\lambda$).  The precise relationship between these approximations and their regimes of validity will be clarified in detail in \cite{aniso}.

\subsection{Temperature Anisotropy}
\label{temp-anisotropy}

As a first step, we compute the CMB temperature anisotropy due to the collision perturbation.  The temperature anisotropy \eqref{3dtemp} gives rise to a gravitational potential perturbation $\Phi$, and the CMB temperature observed in a given direction today is the sum of the intrinsic temperature and gravitational potential at the corresponding point on the sphere defined by the earth's past lightcone at $t=t_{dc}$.   For an adiabatic perturbation the gravitational potential $\Phi_{dc} \sim (-2 / 3)(\delta T/T)_{dc}$ \cite{sw, huwhite} , so the temperature anisotropy observed today is $\delta T/T \sim (-1/2)(\delta T/T)_{dc}$.\footnote{In general there are additional contributions from velocity perturbations, but for features larger than about a degree these are subdominant and we will ignore them.}  Therefore for our purposes we can incorporate the Sachs-Wolfe effect by defining an ``effective temperature'' anisotropy based on \eqref{3dtemp}:
\begin{eqnarray} \label{CMBtemp}
T(\theta, \phi) & =&  T_{dc, eff}( D_{dc}\n)/(1+z_{dc})=T_{0}\left[ 1 + \lambda D_{dc} (\mu - \mu_{c}) \Theta(\mu - \mu_{c}) \right],
\end{eqnarray}
where $\lambda \sim - \tilde \lambda/2$ and
\be \label{Teff}
T_{dc,eff}({\bf x})/(1+z_{dc}) \equiv T_{0}\left[ 1 +  \lambda (x-x_{c}) \Theta(x-x_{c}) \right]
\ee
is the effective temperature perturbation taking the gravitational potential perturbation into account, $D_{dc}=\int_{t_{dc}}^{t_{0}} dt'/a(t')$ is the comoving radius where the past lightcone of the Earth today intersects the decoupling surface, $\n$ is a unit vector pointing from the Earth in the direction $(\theta, \phi)$, $\mu = \cos \theta$, and $\mu_{c} = \cos \theta_{c} = x_{c}/D_{dc}$ is the angular radius of the collision disk in the temperature map.  We have chosen coordinates so that the collision disk is centered at $\theta=0$ ($\mu=1$).

From \eqref{CMBtemp} we see that the effect of the collision is to create a disk on the CMB temperature map, inside of which the temperature depends linearly on the cosine of the angle from the center, and outside of which there is no effect.  The angular radius of that disk is $\theta_{c} = \cos^{-1}(x_{c}/D_{dc})$, and the maximum temperature deviation is $(\Delta T/T)_{\rm max} = \lambda (D_{dc}-x_{c})$.  The disk can either be hot or cold depending on the sign of the parameter $\lambda$; in terms of the microphysics of the collision this depends on the characteristics of the other bubble, the domain wall, and the inflaton (see \cite{wwc2} for a discussion).
With the mapping $\mu_{c} \longrightarrow -x_{T}, ~\lambda(D_{dc}-x_{c}) \longrightarrow M-1$  (the collision disk in this paper is centered at $\theta=0$) the temperature anisotropy \eqref{CMBtemp} coincides with the one discussed in \cite{wwc2}.

\section{Polarization Basics}
\label{sec:set-up-pol}

The CMB is polarized because of the Thomson scattering of CMB photons off electrons \cite{1984ApJ...285L..45B, 1997PhRvD..55.1830Z}. This scattering occurs primarily at redshifts centered at decoupling ($z_{dc} \sim 1100$) and reionization ($z_{re} \sim 10$) as shown in \figref{fig-setup}.  The contribution from decoupling depends primarily on known atomic physics \cite{Seager:1999bc}, but there is currently significant uncertainty in the  reionization history.  We will derive an expression for the contribution to the polarization from the collision in terms of a general visibility function, and then evaluate it using two fiducial reionization models.

As  discussed above in Sec.~\ref{temp-anisotropy}, a bubble collision introduces an inhomogeneity in the temperature at the decoupling surface. As a result, some of the electrons in our past lightcone see a quadrupole temperature anisotropy from their CMB skies. Photons scattered off such electrons then give rise to polarization of the CMB. This section briefly reviews the physics of this effect \cite{1984ApJ...285L..45B, 1997PhRvD..55.1830Z, Hu:1999vq,Dvorkin:2007jp}.

\begin{figure}[t]
\begin{center}
\includegraphics[width=0.6\textwidth]{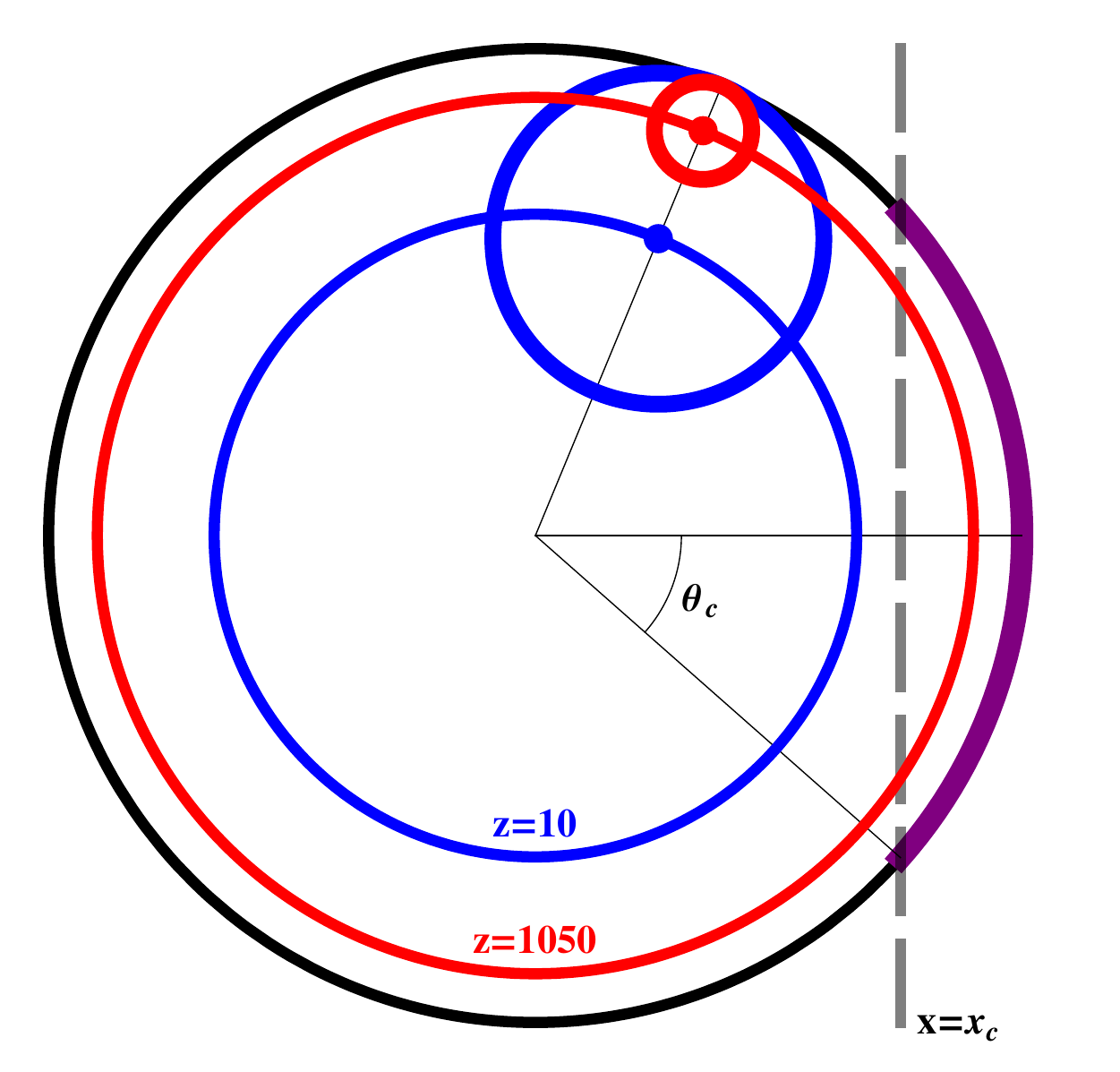}
\end{center}
\caption{\label{fig-lightcones} The geometry of the Earth's past light cone projected on the $xy$ plane of the decoupling surface, showing the primary reionization and decoupling scattering surfaces at $z\sim 10$ and $z\sim 1050$ respectively, and the decoupling sky as seen by an electron at some angle on those surfaces.  The $w$ direction is suppressed. }
\end{figure}

Consider a typical electron in the Universe  at some time $t=t_{e}$ and comoving location ${\bf x}_e$.  Photons scattered by this electron and observed by us today will be polarized if they originate from a temperature distribution such that the anisotropy pattern in the CMB sky \emph{seen by the electron} has a non-zero quadrupole moment.  Therefore as a first step, we should compute the quadrupole moment of the temperature anisotropy on the intersection of the past lightcone of the electron with the decoupling surface.  For an electron at comoving distance $D_{e}$, that intersection is a sphere of comoving radius $R_{e} = D_{dc}-D_{e}$ (see \figref{fig-lightcones}).  Therefore for an electron at position ${\bf x}_e$:
\begin{equation}
T_{2m} ({\bf x}_e) \equiv  \int d\n_e\, Y_{2m} ^* (\n_e)\, \left( \delta T/T \right)_e = {1 \over (1+z_{dc}) T_0} \int d\n_e\, Y_{2m} ^* (\n_e)\, T_{dc, eff}({\bf x_e} + R_{e}\n_e)\Label{quadrupole}
\end{equation}
Here $\n_e$ is a unit vector pointing from the electron at position ${\bf x}_e$ in the direction $(\theta_{e},\phi_{e})$, and $T_{dc, eff}({\bf x_e} + R_{e}\n_e)$ is the effective temperature distribution incident on the electron ({\em c.f.} (\ref{Teff})).  We have used the independence of $\delta T/T$ on redshift and the orthogonality of the monopole and quadrupole in the second equality.

Photons scattered off the electron carry linear polarization proportional to the quadrupole of \eqref{quadrupole}. To find the polarization at a fixed direction $\n$ on our sky, one must integrate the contributions of all electrons located along the line of sight
\begin{equation}
{\bf x}_e = D_{e} \n.
\end{equation}
 The appropriate measure of integration is $g(D)\, dD$, which represents the probability that a CMB photon observed today scattered in the interval $dD$ and has traveled freely since then. Here, $g(D)$ is the visibility function and is related to the optical depth $\tau(D)$ according to
\begin{equation}
g(D) = -\frac{d\tau}{dD}e^{-\tau(D)}. \label{defgd}
\end{equation}
The resulting polarization is \cite{hu, Dvorkin:2007jp}
\begin{equation}
(Q \pm i U) (\n) = { \sqrt{6} \over 10} \int d D\, g(D) \sum_{m=-2} ^{2} T_{2m} (D \n) ~_{\pm 2} Y_{2 m} (\n) \,,\label{qformal}
\end{equation}
where $Q,\,U$ are the Stokes parameters of the CMB polarization and $_{\pm 2} Y_{2 m} (\n)$ are the spin-weighted spherical harmonics. These may be alternatively expressed in terms of the $E$ and $B$ polarization moments
\be
(Q \pm i U) (\n)= - \sum_{lm} ( E_{lm} \pm i B_{lm} ) ~_{\pm 2} Y_{lm} (\n) . \Label{Elms}
\ee

\section{Bubble Collision Polarization}
\Label{calcq}

Armed with \eqref{qformal}, we need $T_{dc, eff}({\bf x_e} + R_{e} \n_{e})$  in order to find the polarization resulting from a bubble collision.  This follows immediately from \eqref{Teff}:
\be \label{deltat}
{ T_{dc, eff}({\bf x_e} + R_{e} \n_{e}) \over (1+z_{dc})T_0} =1 + \lambda R_{e} (\mu_{e} - \mu_{e,c}) \Theta(\mu_{e} - \mu_{e,c}) .
 \ee
Here $\mu_{e} = \cos \theta_{e}= (x-x_{e})/R_{e}$ is the cosine of the angle on the electron's CMB sky, and $\mu_{e,c} = (x_{c}-x_{e})/R_{e}$ is the cosine of the angular radius of the collision disk on the electron's CMB sky ({\em c.f.} \eqref{CMBtemp}).\footnote{If $\mu_{e,c}>1$ the electron does not see the collision, if $\mu_{e,c}<-1$ the electron's entire sky is taken up by it.} 

The effective temperature anisotropy \eqref{deltat} is ready for substitution into the expression for the Stokes parameters.
The planar symmetry of the temperature perturbation $T_{dc}$ means that the only non-vanishing component of the quadrupole moment is
\begin{eqnarray}
 T_{20}({\bf x}_{e}=D_{e} \n) =  {\sqrt{5 \pi} \over 8} \lambda R_{e}
 \left\{1-\left( { D_{dc} \mu_{c} - D_{e} \mu \over R_{e} } \right)^{2} \right\}^{2}  \Theta\left(1-\left( { D_{dc} \mu_{c} - D_{e} \mu \over R_{e} } \right)^{2} \right)  \nonumber \\
   = {\sqrt{5 \pi} \over 8} \lambda (1-\alpha) D_{dc}
 \left\{1-\left( { \mu_{c} - \alpha \mu \over 1-\alpha } \right)^{2} \right\}^{2}  \Theta \left\{1-\left( { \mu_{c} - \alpha \mu \over 1-\alpha } \right)^{2} \right\},
\end{eqnarray}
where in the first equality we have used $R_{e} \mu_{e,c} = x_{c}-x_{e}=D_{dc}\mu_{c}-D_{e}\mu$, and $\alpha \equiv D_{e}/D_{dc}$ is the ratio of the distance to the electron to the distance to the decoupling surface.  The Stokes parameter $U$ vanishes in these coordinates, again due to the planar symmetry, as do all the spin-weighted spherical moments of $Q$ except those with $m=0$.  Therefore $B_{lm}=0$ and the polarization is purely $E$-mode --- as expected for a scalar perturbation.  Using
\begin{equation}
\phantom{f}_{\pm 2}Y_{20}(\mu) = \frac{3}{4} \sqrt{\frac{5}{6\pi}} (1-\mu^2)\,,
\end{equation}
we find
\begin{equation}  \Label{genstokes}
Q(\theta) = \frac{3}{64}\left({\Delta T \over T_{0} }\right)_{\rm max} {1-\mu^2 \over 1-\mu_{c}} \int_{0}^1 d\alpha D_{dc} \,g\left(\alpha D_{dc}\right)
(1-\alpha)  \left\{1-\left( { \mu_{c} - \alpha \mu \over 1-\alpha } \right)^{2} \right\}^{2}   \Theta(\ldots),
\end{equation}
where the argument of the step function is the quantity in $\{\,\}$, and $\left({\Delta T /T_{0} }\right)_{\rm max} = \lambda (D_{dc}-x_{c})$ is the temperature perturbation at the center of the collision disk.  For a hot spot $\left({\Delta T / T_{0} }\right)_{\rm max}>0$ and $Q<0$ (recall that $g= -({d\tau}/{dD})e^{-\tau}<0$), while for a cold spot $Q>0$.

\section{Analysis of $Q(\theta)$} \label{sec-analytic}
With formula \eqref{genstokes} in hand, our goal is to compute $Q(\theta)$. In \eqref{genstokes},  $Q(\theta)$ is determined up to the behavior of $g(\alpha D_{dc})$, the visibility function in \eqref{defgd} that encodes the information about when CMB photons  last scattered.  It is normalized by
\be \label{visnorm}
\int_{D_{dc}} ^0 g(D) dD= - \int _0 ^1 d\alpha \, D_{dc} \, g(\alpha D_{dc} )=1\,.
\ee
Photons are most likely to scatter at roughly two times: within one scattering length of decoupling (at a redshift of approximately $z \sim 1050$), and at reionization ($z \sim 10$).

The atomic physics at decoupling is well understood, and we can compute the visibility function there to good accuracy as a function of the cosmological parameters using \cite{Seager:1999bc}.  However, the approximations we used in computing the temperature anisotropy from the collision break down at angular scales of about a degree and below, due to effects such as  acoustic oscillations in the plasma between reheating and decoupling, Silk damping, the finite thickness of the decoupling surface, etc.  We are currently engaged in studying these effects in detail \cite{aniso}.  For now, one must bear in mind that the signatures presented here receive  ${\cal O}(1)$ corrections on degree scales and below ($\ell \gtrsim 200$).  This is particularly relevant for the contribution to polarization due to scattering at decoupling, since its details depend on the sub-degree structure of the temperature distribution.  Nevertheless, the rough ($>1^{{\circ}}$ scale) features of the signature we calculate are robust to these corrections.

The physics responsible for reionization is less well understood, and our results are somewhat sensitive to the reionization model. Therefore we will proceed as far as possible with a model-independent analysis, and then study some representative reionization models. Using some reasonable assumptions we will see that we can make good estimates for $Q(\theta)$ and quantify the expected effects from a collision into a unique signal that can be correlated with the effect on the CMB temperature map from the collision, e.g. the WMAP cold spot \cite{coldspot1,coldspot2,coldspot3,coldspot4,coldspot5}.

\subsection{The Model-Independent Geometric Factor}
To isolate the dependence on the visibility function we rewrite \eqref{genstokes} as
\beq \label{K}
Q(\theta) &=&  \int_{0} ^1 d\alpha \, D_{dc}   g(D_{dc} \, \alpha)  \\
&\times& \underbrace{ \left\{  \frac{3}{64}\left({\Delta T \over T_{0} }\right)_{\rm max} {1-\mu^2 \over 1-\mu_{c}} (1-\alpha)  \left\{1-\left( { \mu_{c} - \alpha \mu \over 1-\alpha } \right)^{2} \right\}^{2}   \Theta(\ldots)\right\} }_{\equiv K( \alpha, \theta)} ~,\nonumber
\eeq
such that $K( \alpha,\theta)$ is dimensionless and contains all the recombination and reionization model-independent factors.

\subsubsection{Analysis of $K(\alpha,\theta)$} The function $K(\alpha,\theta)$ encodes the contribution of the collision perturbation to the Stokes paramater $Q(\theta)$ due to Thomson scattering by a surface of electrons in our past lightcone at a comoving distance $\alpha D_{dc}$. This contribution is weighted by the visibility function to give the total $Q$, which is dependent on the model for recombination and reionization. It is thus instructive to examine the behavior of $K$ in more detail.

Thomson scattering at $\alpha D_{dc}$  will lead to an {\em annulus} of $Q$-mode polarization on the sky, centered at $\theta=0$ (the center of the spot).  The inner radius of the annulus may be either zero (making it a disk) or non-zero, and its outer radius is always greater than $\theta_{c}$.  By symmetry one can see that the polarization vanishes at $\theta=0$, which is encapsulated by the geometric prefactor $(1-\mu^2)$.

The radii of the annulus are determined by the step function in \eqref{K}. The argument of the step function is non-vanishing when
\be
{\mu_c \over \alpha} - {1-\alpha \over \alpha} < \mu < {\mu_c \over \alpha} + {1-\alpha \over \alpha}   ~.\Label{bounds}
\ee
For there to be a annulus rather than a disk, the upper bound in \eqref{bounds} must be $<\!1$. Physically, this corresponds to a region on the CMB sky for which all points on the electron's decoupling sphere are inside the collision region $x>x_{c}$.  In this case the electron sees a pure dipole on its sky, so there is no quadrupole moment to contribute to polarization. In \figref{fig-lightcones} this happens when the circles corresponding to the electrons' past light cones are small enough, and the angle close enough to $\theta=0$, that they fit entirely in the region to the right of $x=x_c$.
In terms of the inequality in \eqref{bounds} this occurs when 
\be
\mu_c < 2 \alpha -1 ~.
\ee
When this is satisfied the inner radius is
\be
\theta_{\rm inner}=\cos^{-1} \left( {\mu_c \over \alpha} + {1-\alpha \over \alpha}\right) \leq \theta_{c},
\ee
and in all cases the outer radius is
\be
\theta_{\rm outer}=\cos^{-1} \left( {\mu_c \over \alpha} - {1-\alpha \over \alpha}\right) \geq \theta_{c},
\ee
where the inequalities are saturated only if $\alpha=1$ (the scattering electron is at the decoupling surface).

\subsection{Analysis of the Stokes Parameter} \Label{sec-results}
We have analyzed the model-indepedent geometric factor involved in computing to the Stokes parameter. In this section, we will employ some fiducial models for the ionization fraction from reionization (combined with the well-understood ionization fraction from recombination) to compute the resulting Stokes parameter. The first step is to compute the visibility function.

\subsubsection{The Visibility Function}  Recall that the visibility function as a function of redshift is
\be
g(z) = -{d\tau \over dz} e^{-\tau(z)} ~,
\ee
where $\tau(z)$ is the optical depth, and $g(z)$ is normalized so that $-\int_0^\infty dz~g(z)=1$. The optical depth to scattering out to cosmic time $t$ is 
\be
\tau(t) = \int_t ^{t_0} \sigma_T \,  n_e(t) dt ~,
\ee
where $\sigma_T$ is the Thomson cross-section, $n_e$ is the number density of \emph{free} electrons and $t_0$ is the age of the Universe. In terms of redshift this becomes 
\be
\tau(z)=n^{\rm tot}_{p,0} \sigma_T \int_0^z {(1+z')^2 \over H} \chi_e (z') dz' ~,
\ee
where $n^{\rm tot}_{p,0}$ is the total number density of protons today (both bound and free), $\chi_e\equiv n_e/n^{\rm tot}_{p,0}$ is the ionization fraction, and $
H(z) \equiv H_0 \sqrt{ \Omega_r (1+z)^4 +\Omega_m (1+z)^3+\Omega_k(1+z)^2 + \Omega_\Lambda }
$
is the Hubble rate at redshift $z$.
Here, $H_0$ is the present Hubble rate, and the $\Omega_{x}$ (for $x \in \{r,m,k,\Lambda\}$) are the fractional energy density of the Universe in radiation, matter, curvature and dark energy (cosmological constant) respectively.  We assume a standard helium fraction of $Y_{He} \simeq 0.24$ and use the best fit values from \cite{Komatsu:2010fb} as reference cosmological parameters.  Thus, our only other input is the ionization fraction, which characterizes the physics of decoupling and reionization.  For a given ionization history it is straightforward to compute the visibility function.  It is physically illuminating to trade $z$ for the comoving distance $D$ as the independent parameter in the visibility function using $g(D) dD = g(z) dz$ where
\be
D(z) =  \int_{(1+z)^{-1}}^{1} {da \over a^2 H} = {1 \over H_0} \int_{(1+z)^{-1}} ^1 d {a} f( a) ~, 
\ee
and
$f(a) = [\Omega_r + \Omega_m a +\Omega_k a^2 + \Omega_\Lambda a^4]^{-1/2} $.
The physics of decoupling is well understood, and we compute the ionization fraction using \cite{Seager:1999bc}. For reionization, we will take two representative models, the  single reionization paramatrization used by CAMB \cite{Lewis:2008wr}, and a more complex double reionization model to qualitatively span the likely space of possible reionization histories \cite{Mortonson:2007hq}.

\begin{figure}[t]
\begin{center}$
\begin{array}{cc}
\includegraphics[width=0.5\textwidth]{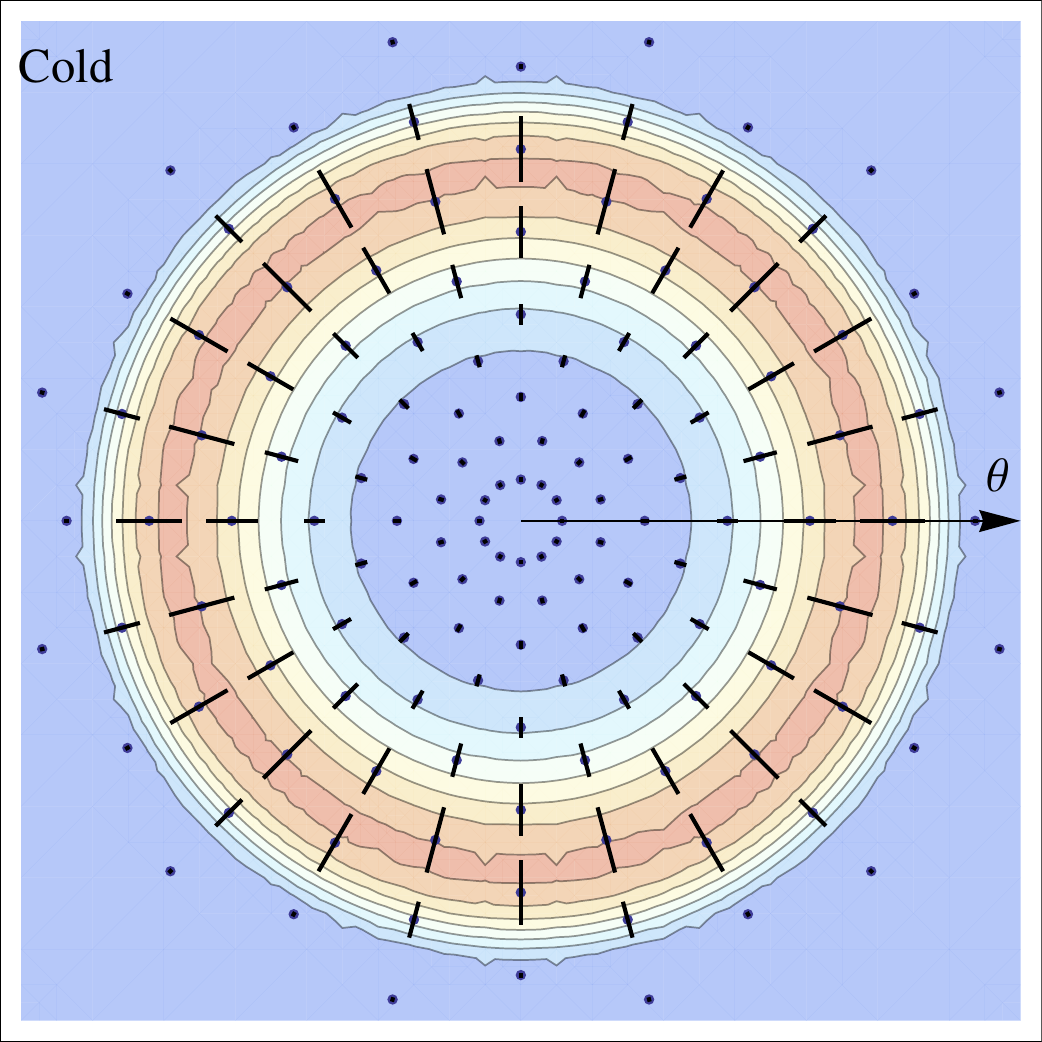}
\includegraphics[width=0.5\textwidth]{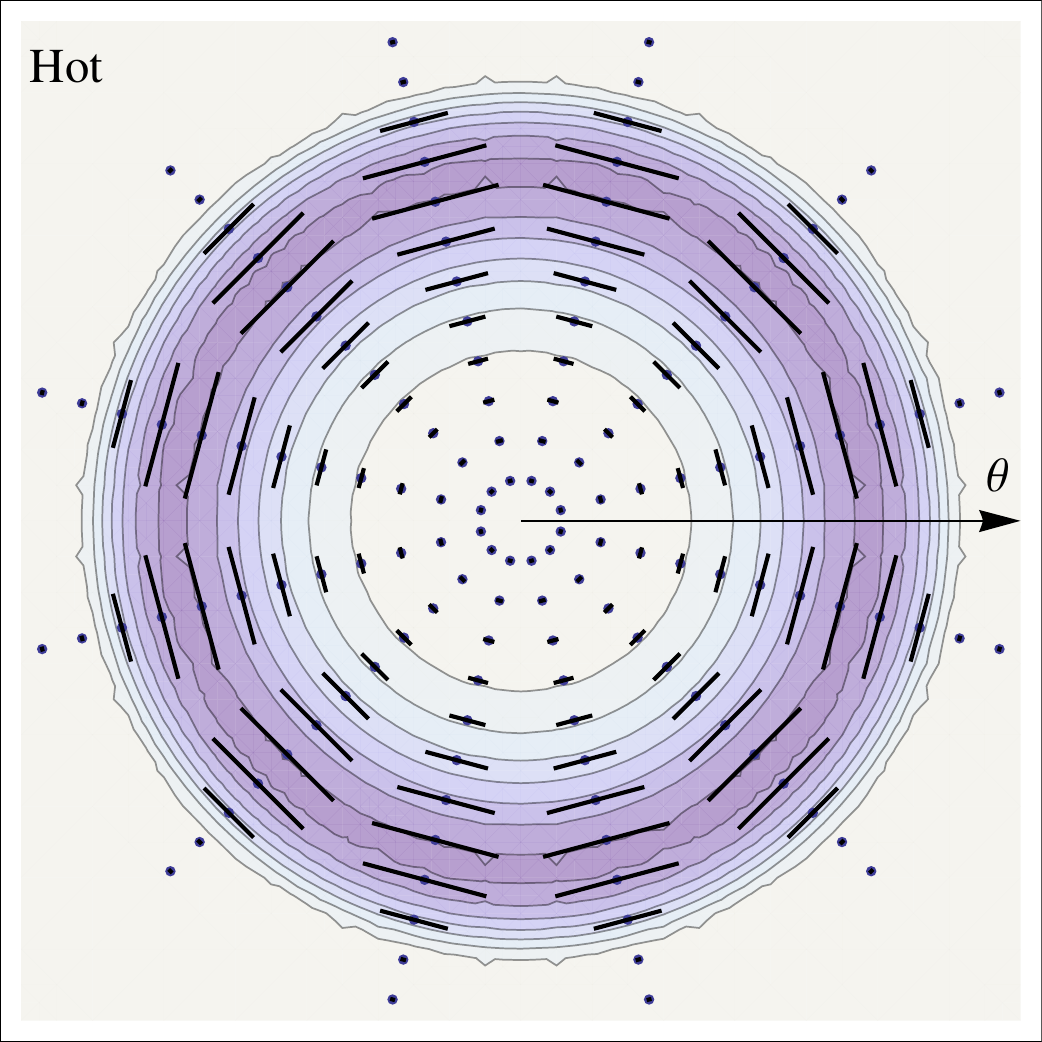}
\end{array}$
\end{center}
\caption{\label{fig-polpatterns} Polarization patterns around cold ($Q>0$) and hot ($Q<0$) spots.}
\end{figure}

\subsubsection{Single Reionization} 

\begin{figure}[t]
\begin{center}$
\begin{array}{cc}
\includegraphics[width=0.4\textwidth]{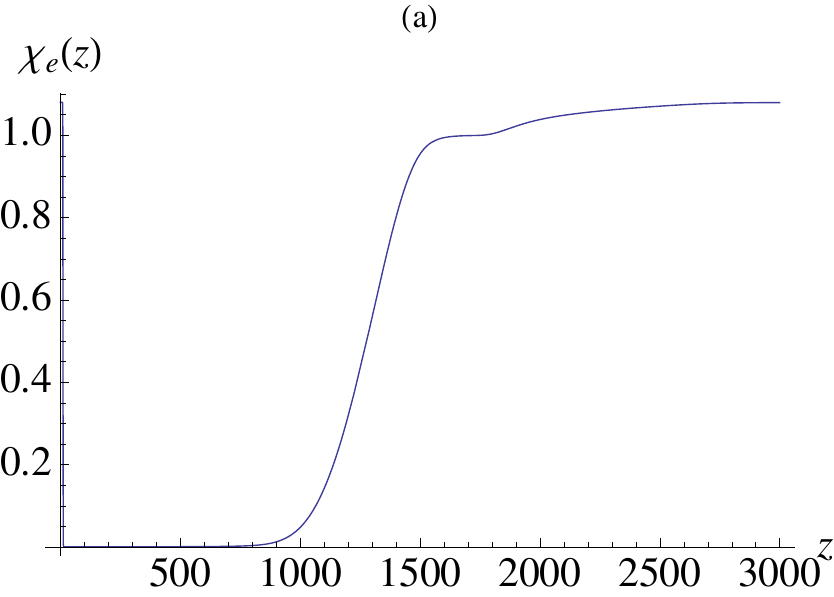}
\includegraphics[width=0.4\textwidth]{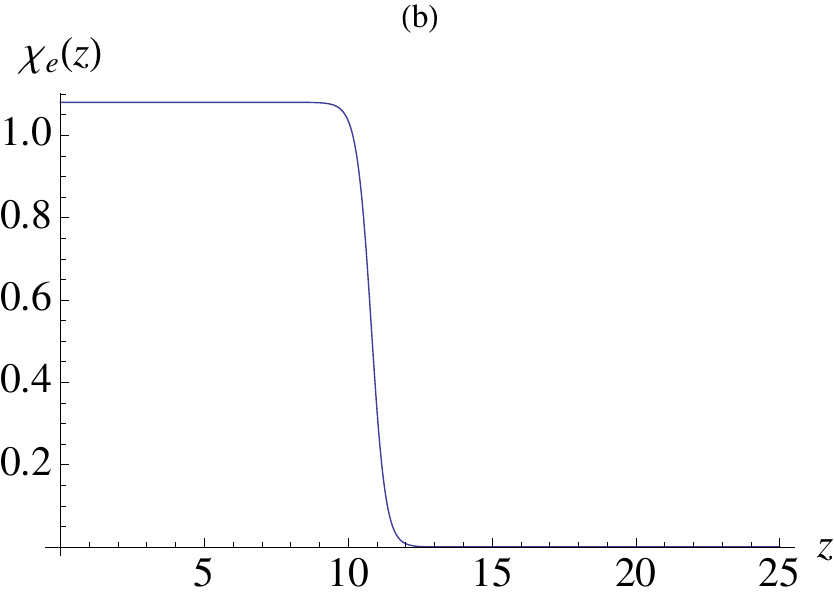}
\end{array}$
\end{center}
\caption{\label{fig-ionfrac1} The ionization fraction in the single reionization model with $z_{re} = 10.8$,  $\Delta_z = 0.5$ and $h=1.08$. We normalize such that $\tau_{re} = 0.09$; the early time ionization fraction is computed using RECFAST with best fit values from WMAP-7 year data \cite{Komatsu:2010fb}. Figure (a) is for all times and (b) zooms in on the reionization epoch.}
\end{figure}

We take the ionization fraction at late times to have the form
\beq
\chi_{re} (z) &=& {h \over 2} \left[ 1 - \tanh\left( {u-u(z_{re} ) \over \Delta_u } \right) \right] ~, \\
u(z) &=& (1+z)^{3/2} ~, \qquad \Delta_u = {3 \over 2} \sqrt{1+z_{re} } \Delta_z ~, \nonumber
\eeq
where $h$, $z_{re}$ and $\Delta_z$ are input parameters.  We set $h=1.08$ in agreement with CAMB (taking the first reionization of helium into account) \cite{Lewis:2008wr}. We choose values that produce a realistic ionization fraction at reionization \cite{Lewis:2008wr}, normalize such that $\tau_{re}=0.09$, $z_{re}=10.8$ are in agreement with WMAP-7 year data, and choose $\Delta_z=0.5$. The total ionization fraction is $\chi_e (z) = \chi_{re} (z) + \chi_{dc} (z)$ where $\chi_{dc}$ is taken from RECFAST \cite{Seager:1999bc} with best fit values from the WMAP-7 year data \cite{Komatsu:2010fb}.  We plot the ionization history for this choice in \figref{fig-ionfrac1}. Equipped with this, we compute the visibility function $g(\alpha D_{dc})$ using the steps outlined above (taking $D_{dc} = D(z=2000)$ as our reference distance scale), and then use \eqref{genstokes} to compute the Stokes parameter $Q$. 

\begin{figure}[t]
\begin{center}$
\begin{array}{cc}
\includegraphics[width=0.4\textwidth]{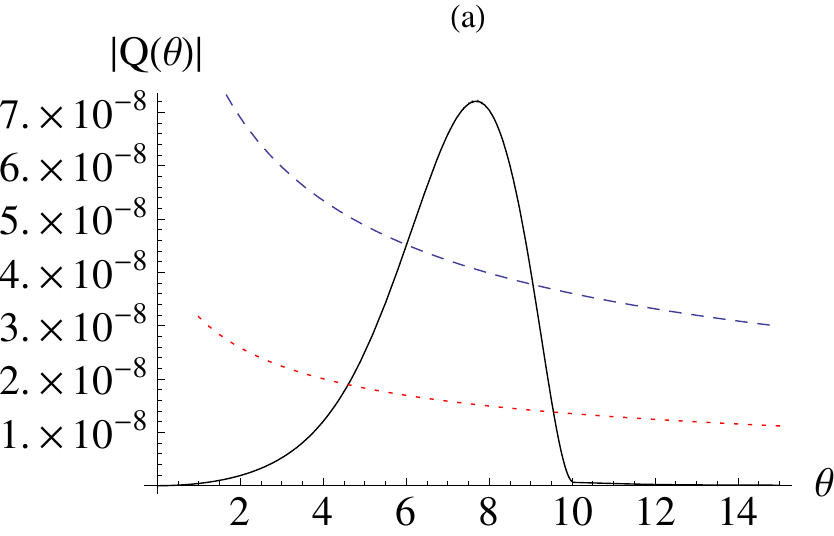}
\includegraphics[width=0.4\textwidth]{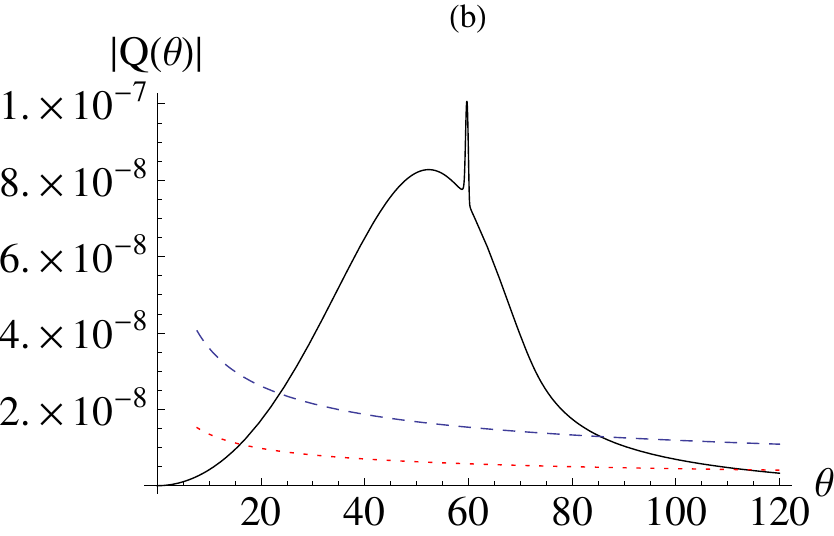}
\end{array}$
\end{center}
\caption{\label{fig-singlestokes} The Stokes parameter $|Q(\theta)|$ vs. $\theta$ (solid black line) for the single reionization model of \figref{fig-ionfrac1}. Figure (a) is the small spot $|\Delta T / T_0|_\textrm{max} = 2 \times 10^{-4}$ and an angular radius of $10^\circ$. The dominant contribution for the small spot comes from scattering at decoupling. Figure (b) is the large spot with $|\Delta T / T_0|_\textrm{max} = 5 \times 10^{-5}$ and an angular radius of $60^\circ$. For the large spot the sharp feature is the contribution from scattering at decoupling, the rest of the function comes from scattering after reionization. $Q<0$ for a hot spot and $Q>0$ for a cold spot, see \figref{fig-polpatterns}. On each we display the estimated sensitivity for both the SPIDER (red dotted line) and Planck (blue dashed line) experiments, averaged over an annulus with inner angular radius $\theta$ and a width of $2^\circ$. }
\end{figure}

We compute $Q$ for a small spot with parameters similar to the WMAP cold spot ($|\Delta T / T_0|_\textrm{max}=2 \times 10^{-4}$ and an angular radius of $10^\circ$), and for a larger spot ($|\Delta T / T_0|_\textrm{max}|=5 \times 10^{-5}$ and angular radius of $60^\circ$). We show the results for $|Q(\theta)|$ in \figref{fig-singlestokes}. The polarization patterns are radial for cold spots ($Q>0$), and tangential for hot spots ($Q<0$), see \figref{fig-polpatterns}. For the small spot, the dominant contribution comes from scattering at decoupling and so is {\it not} sensitive to the physics at reionization. In contrast to this, for the large spot the dominant contribution comes from scattering at reionization, while scattering at decoupling leads to the narrow spike.\footnote{We emphasize again that sub-degree features in our analysis are subject to ${\cal O}(1)$ corrections, because we have not taken the relevant sub-horizon physics into account.  We expect this spike to broaden to roughly a degree, and potentially to develop substructure \cite{aniso}.  However, the qualitative feature---a sharp spike at the edge of the temperature disk---should remain.}   Thus larger spots {\it are} sensitive to the physics at reionization. The Stokes parameters are robust under reasonable variations of the reionization history (within the single reionization model).

To give the reader an estimate of the detectability of our signal we estimate the sensitivity level for two polarization experiments, SPIDER and Planck. We estimate the sensitivity by computing the number of pixels in an annulus with inner radius given by $\theta$ in \figref{fig-singlestokes}(a) and (b) and an angular width of $2^\circ$. On each figure for the Stokes parameter we display the resulting sensitivity level based on current estimates for SPIDER (red dotted line) in the 145 GHz band \cite{spider} and Planck (blue dashed line) in the 143 GHz band \cite{planck2}. For SPIDER we assumed a 25 day flight and and coverage of roughly 50\% of the sky (which includes the area of the cold spot). For Planck the data is over the full expected 15 month lifetime of the instrument. For both the small and large spot the signal should be detectable based on current estimates of noise levels and detector sensitivity.

\begin{figure}[t]
\begin{center}$
\begin{array}{cc}
\includegraphics[width=0.4\textwidth]{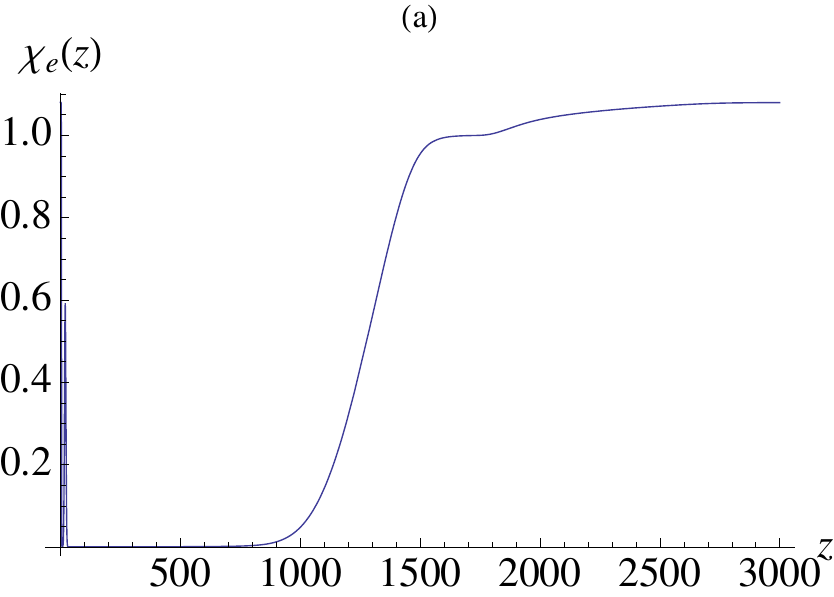}
\includegraphics[width=0.4\textwidth]{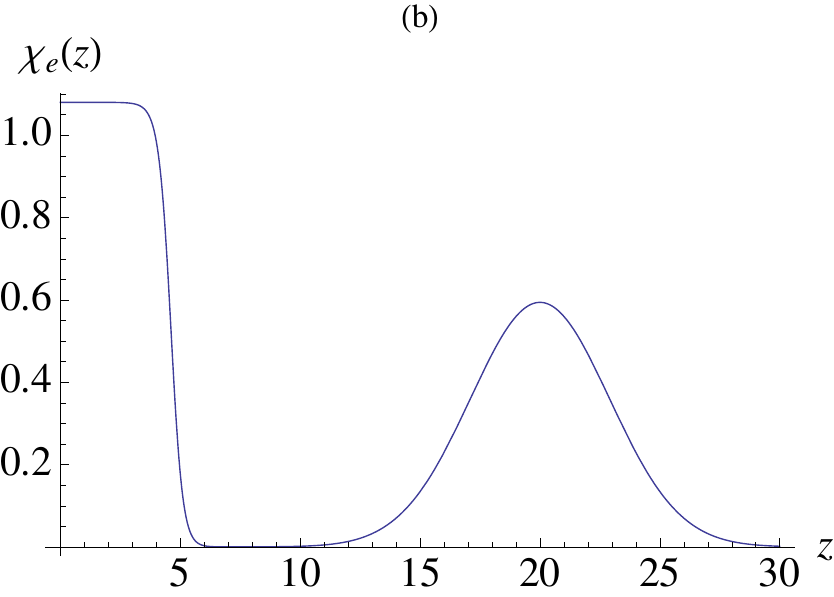}
\end{array}$
\end{center}
\caption{\label{fig-ionfrac2} Ionization fraction in the double reionization model with  $\tilde h /h =1.1$, $z_1=4.6$, $\Delta_z =0.5$, $z_2=20$ and $\sigma_2=2.9$. We normalize such that $h=1.08$ and $\tau_{re} = 0.09$ in agreement with WMAP-7 year data. The early time ionization fraction is computed using RECFAST with best fit values from WMAP-7 year data \cite{Komatsu:2010fb}. Figure (a) is for all times and (b) zooms in on the reionization epoch.}
\end{figure}

\begin{figure}[t]
\begin{center}$
\begin{array}{cc}
\includegraphics[width=0.4\textwidth]{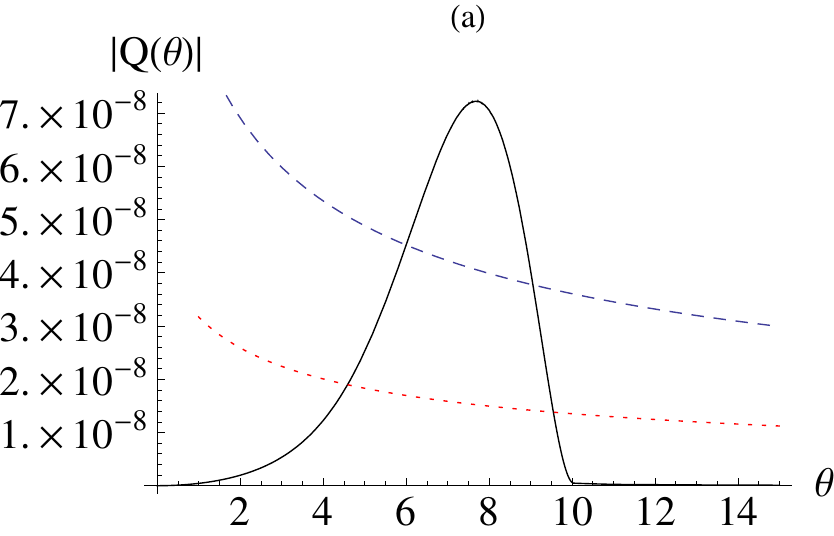}
\includegraphics[width=0.4\textwidth]{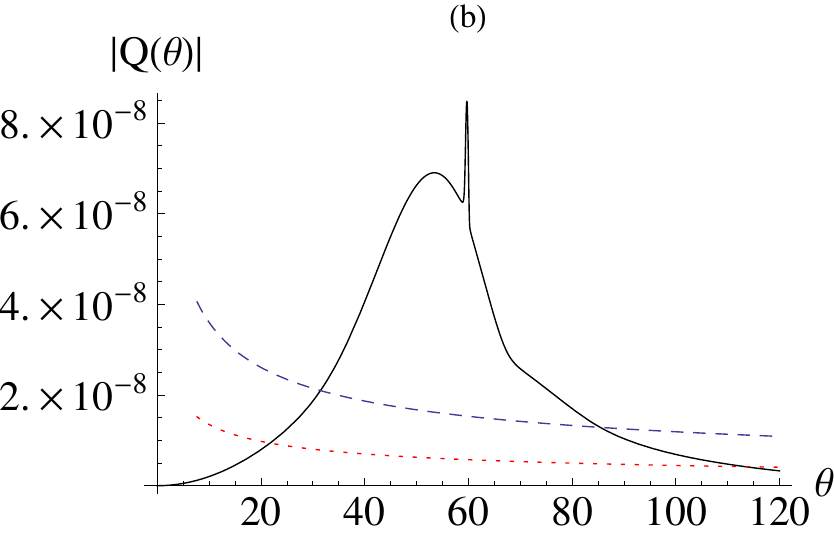}
\end{array}$
\end{center}
\caption{\label{fig-doublestokes} The Stokes parameter $|Q(\theta)|$ vs. $\theta$ for the double reionization model of \figref{fig-ionfrac2}. Figure (a) is the small spot $|\Delta T / T_0|_\textrm{max} = 2 \times 10^{-4}$ and an angular radius of $10^\circ$. The dominant contribution for the small spot comes from scattering at decoupling. Figure (b) is the large spot with $|\Delta T / T_0|_\textrm{max} = 5 \times 10^{-5}$ and an angular radius of $60^\circ$. For the large spot the sharp feature is the contribution from scattering at decoupling, the rest of the function comes from scattering after reionization. $Q<0$ for a hot spot and $Q>0$ for a cold spot, see \figref{fig-polpatterns}. On each we display the estimated sensitivity for both the SPIDER (red dotted line) and Planck (blue dashed line) experiments, averaged over an annulus with inner angular radius $\theta$ and a width of $2^\circ$.}
\end{figure}

\subsubsection{Double Reionization} We model double reionization using
\be
\chi_{re} (z) = {h \over 2} \left[ 1 - \tanh\left( {u-u(z_{1} ) \over \Delta_u } \right) \right] +{ \tilde h \over 2} e^{-(z-z_2)^2 \over 2 \sigma_2 ^2} ~,
\ee
where  $h$, $z_{re}$, $\Delta_z$, $\tilde h$, $z_2$ and $\sigma_2$ are input parameters. We again normalize using $h=1.08$. The other parameters are free, but must be chosen such that $\tau_{re}=0.09$. As a typical example, we choose $\tilde h /h =1.1$, $z_1=4.6$, $\Delta_z =0.5$, $z_2=20$ and $\sigma_2=2.9$. The ionization fraction is shown in \figref{fig-ionfrac2}. This function is a good approximation to that of \cite{Mortonson:2007hq}. We compute the Stokes parameter $Q$ using the same spot parameters as before, and display this in \figref{fig-doublestokes}.

As in the single reionization case, we see that for small spots the dominant contribution is from scattering at decoupling, while for large spots it is scattering at reionization. The Stokes parameters are again robust under reasonable variations of the reionization parameters within the double reionization model. For the large spot, $Q$ differs slightly from the single reionization model. We again display the expected sensitivities of SPIDER in the 143 GHz band (red dotted line) and Planck in the 145 GHz band (blue dashed line) on each figure.

The polarization pattern for the models plotted in \figref{fig-singlestokes} and \figref{fig-doublestokes} would be detectable by both Planck and SPIDER given their instrumental sensitivities and angular resolutions.  A smaller-scale polarization survey, with better resolution and sensitivity and focused on the cold spot, could improve on this.  

\section{Differentiability from Primordial Fluctuations} \label{seccorr}

\begin{figure}[t]
\begin{center}$
\begin{array}{ccc|ccc}
\includegraphics[width=0.33\textwidth]{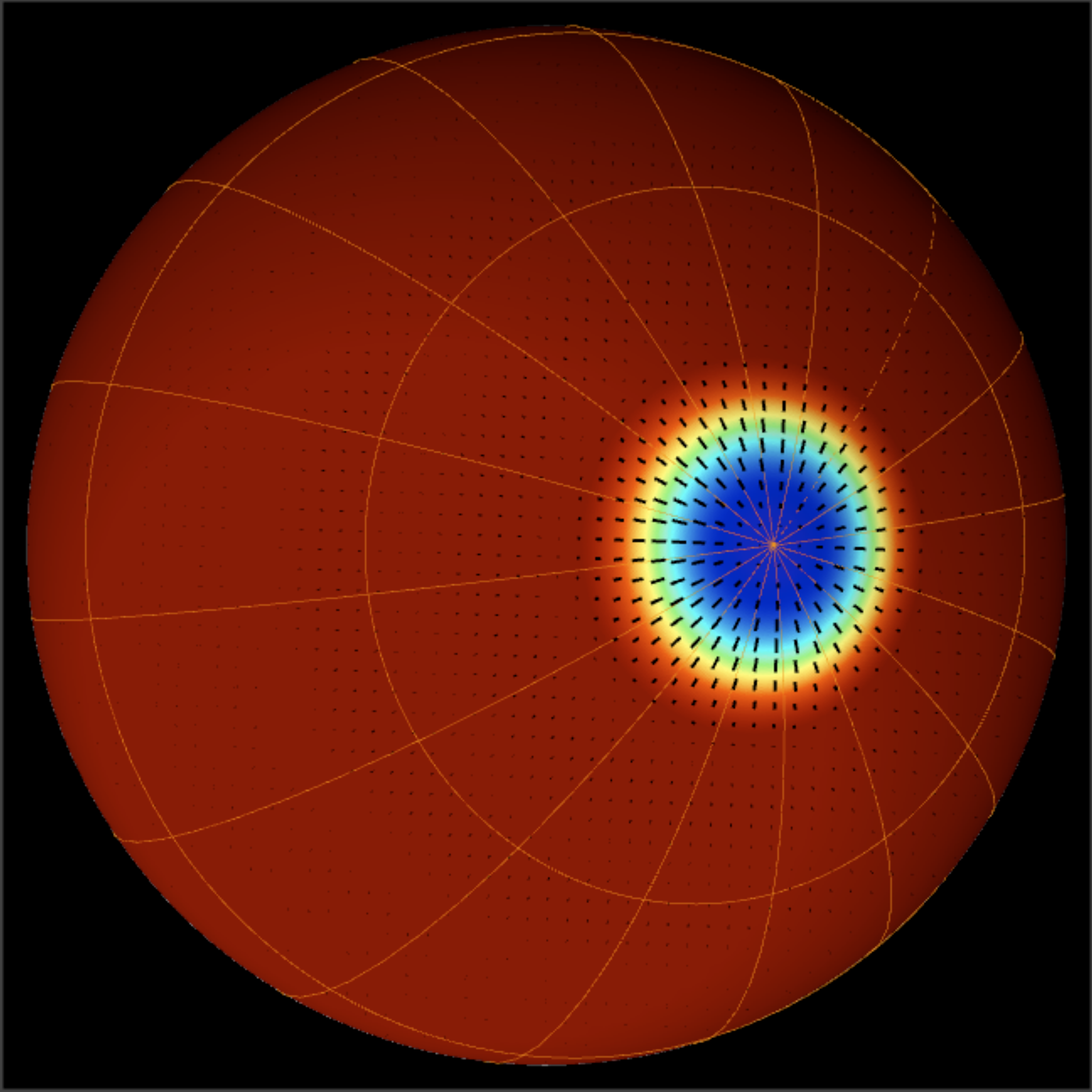}
\includegraphics[width=0.33\textwidth]{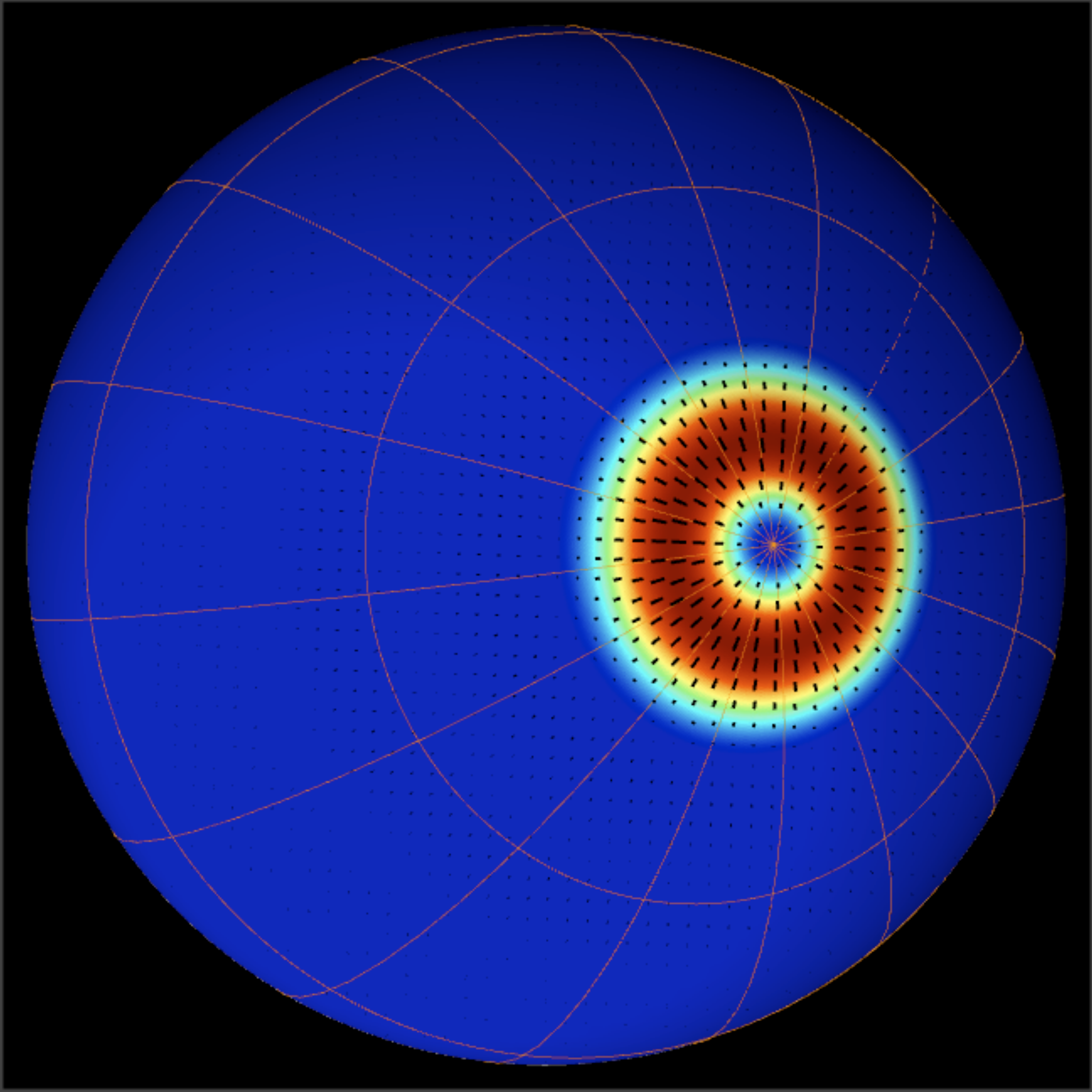}
\includegraphics[width=0.33\textwidth]{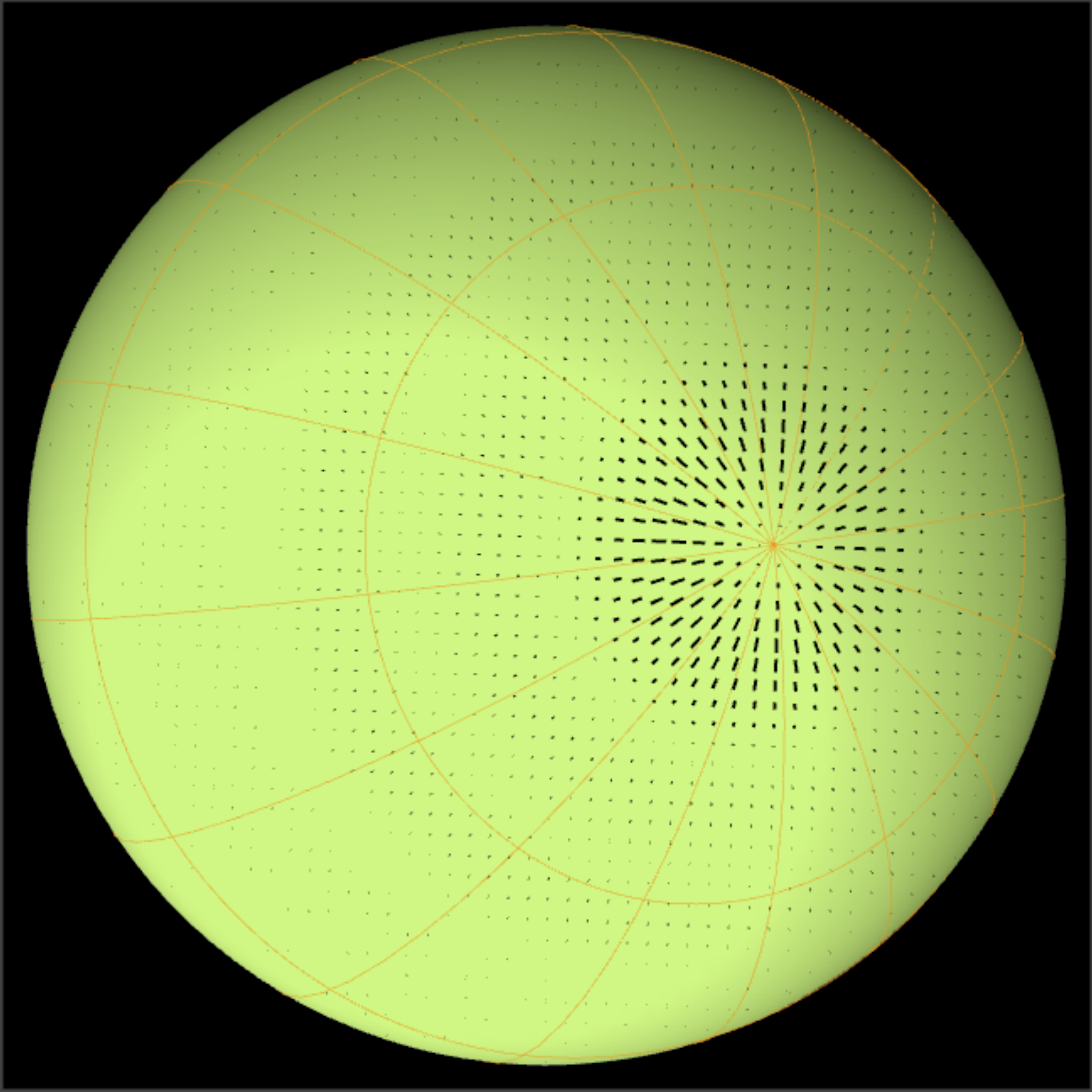} \\
\includegraphics[width=0.33\textwidth]{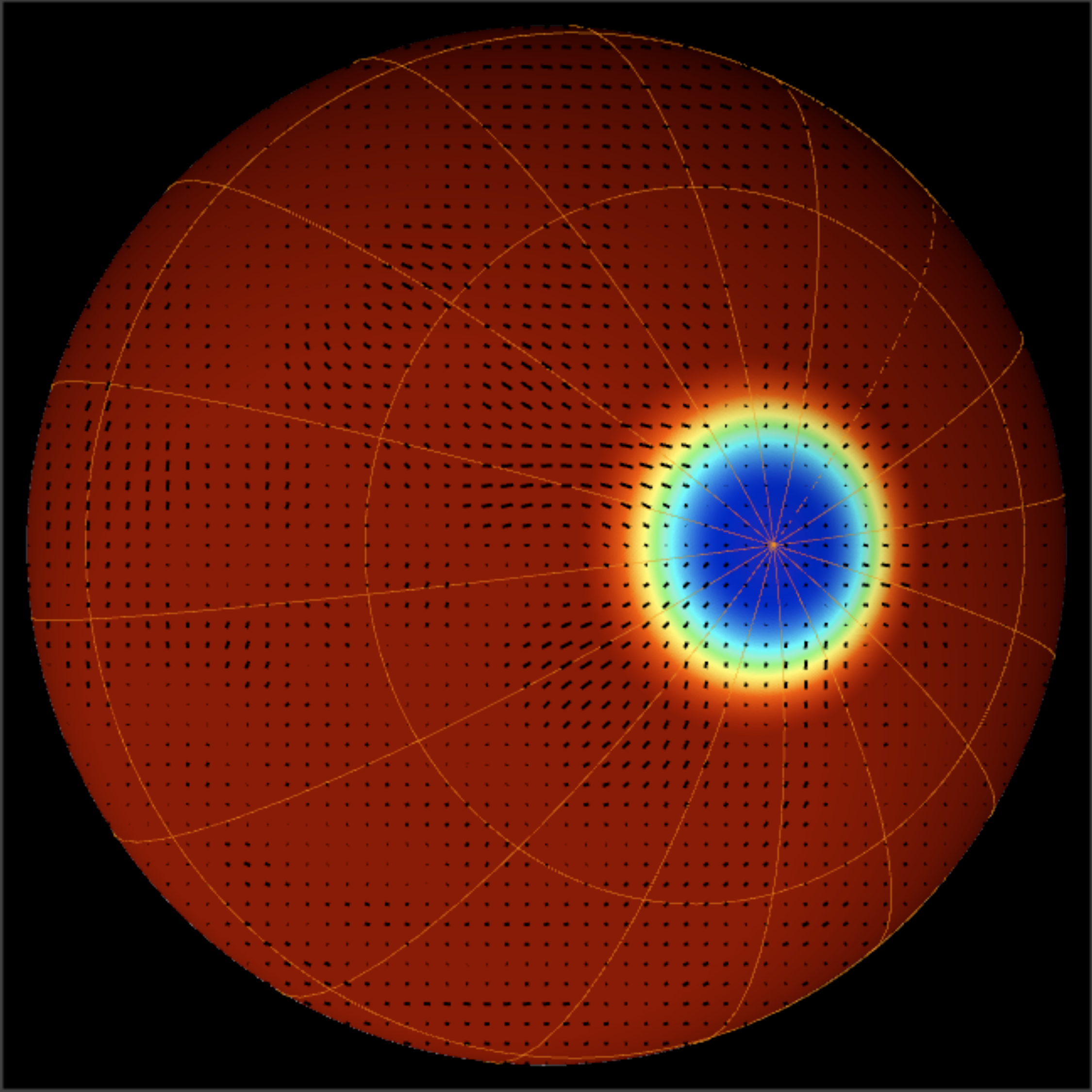}
\includegraphics[width=0.33\textwidth]{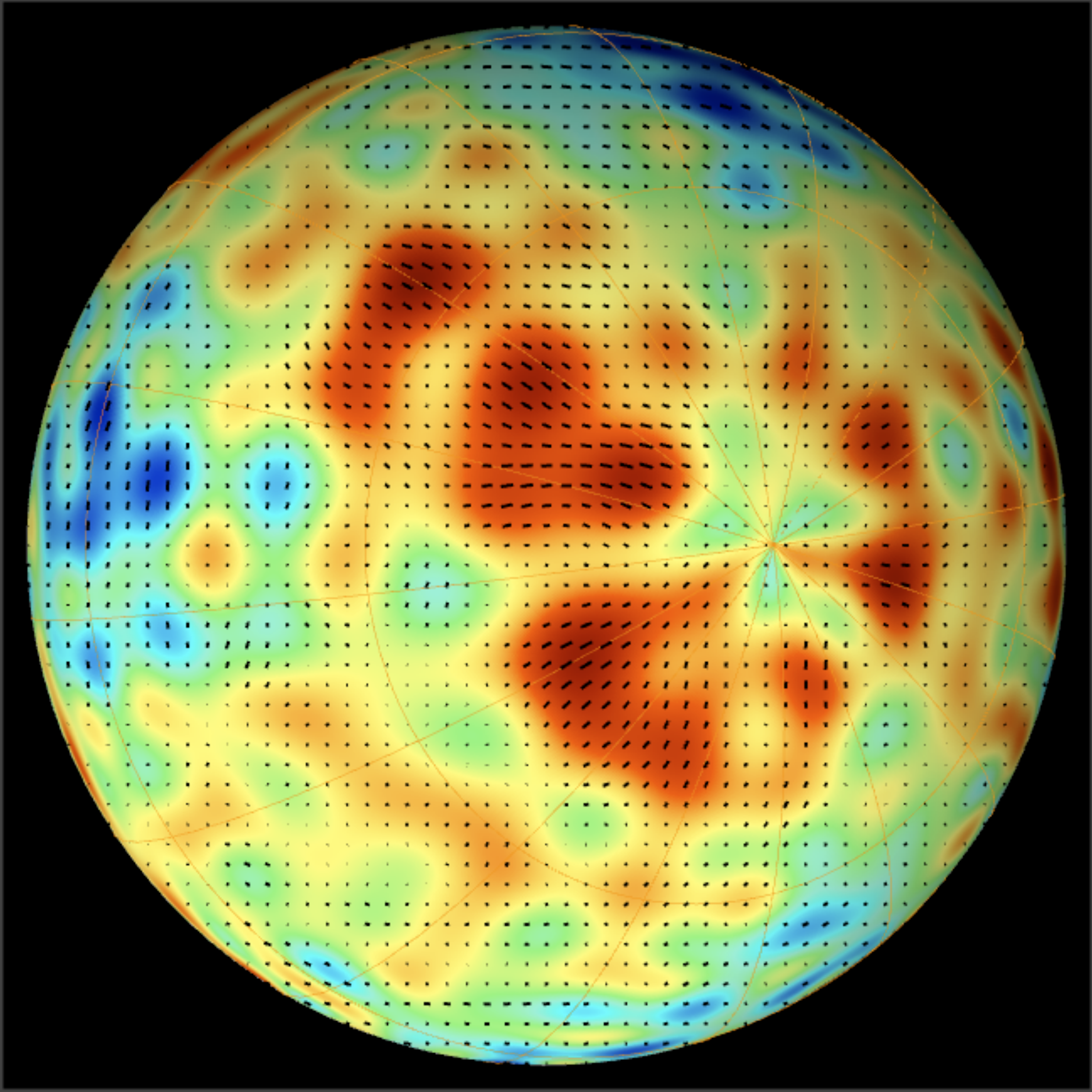}
\includegraphics[width=0.33\textwidth]{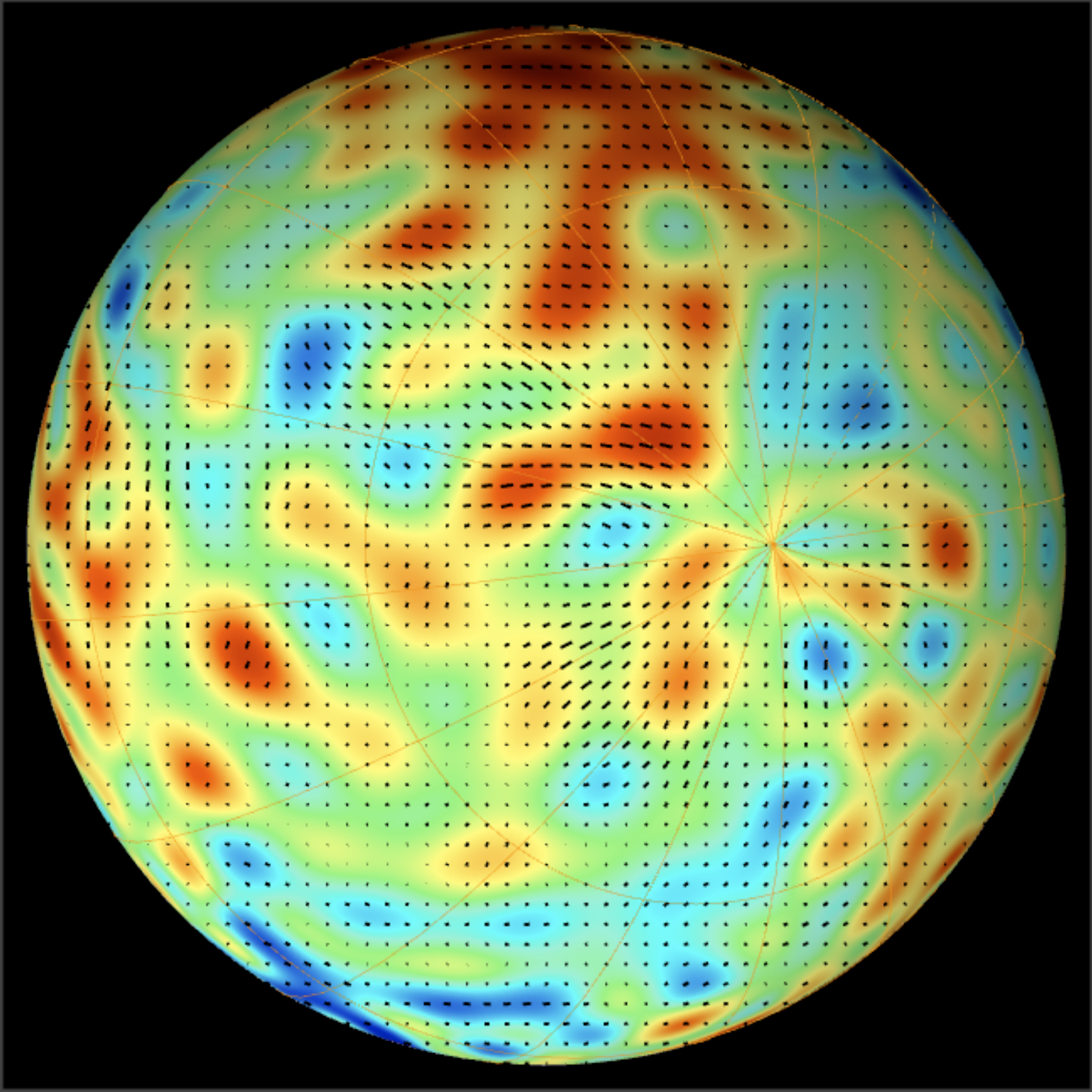}
\end{array}$
\end{center}
\caption{\label{fig-tqu} The upper panels are (from left to right) the circularly-symmetric T, Q, and U patterns resulting from the three-dimensional planar symmetry of a bubble-collision generated cold spot with an angular radius of 10 degrees and a peak temperature deficit $\Delta T/T = -2 \times 10^{-4}$.  The lower left panel is an identical circularly-symmetric T pattern but the Q and U panels are now a realization of a Gaussian random field statistically consistent with the T pattern, and the polarization power spectra $C_{l}^{EE}$ and $C_{l}^{TE}$.  This illustrates that generic polarization patterns consistent with a two-dimensional cold spot need not have the full three-dimensional planar symmetry expected from a bubble collision.}
\end{figure}

The temperature and polarization anisotropies generated in the CMB from a bubble collision reflect an underlying planar symmetry arising from the direction picked out by the collision.   This planar symmetry differentiates the polarization pattern generated via a bubble collision from those due to generic random Gaussian fluctuations.    This is qualitatively illustrated in \figref{fig-tqu} where the temperature and polarization anisotropies from a bubble collision are contrasted with those from a Gaussian random field constrained to produce a temperature anisotropy identical to that from a bubble collision and a polarization pattern consistent with the angular power spectra $C_l^{EE}$ and $C_l^{TE}$.  For a Gaussian random field it is not necessary for the polarization pattern to be circularly-symmetric even if the temperature pattern is, and this reflects the fact that the projection from three-dimensional curvature perturbations to two-dimensional temperature perturbations is not unique --- many three-dimensional patterns can produce the same two-dimensional pattern.   
Randomly generating this planar symmetry would be highly unusual, and so it is clear that a highly correlated and symmetric temperature and polarization pattern represents a non-trivial test of a bubble-collision origin for features in the CMB.   We leave the important issue of quantifying the expected signal-to-noise of these correlated polarization features and a detailed analysis of detectability versus random fluctuations to future work \cite{aniso}.

\section{Conclusions}

In this paper we have computed the polarization of CMB photons due to a cosmic bubble collision. Together with the results of \cite{wwc2}, these results constitute a specific, quantitative prediction for the effects of a cosmic bubble collision on the CMB.  However, there are other models that could produce signals of this type.  Chief among them is of course concordance inflationary cosmology, which predicts the existence of random Gaussian fluctuations in the temperature map, along with their associated polarizations.  Can one distinguish the effects of a bubble collision from those of an unusually large primordial random fluctuation?

To answer this question in detail would require an analysis beyond the scope of this paper.  However there is a sharp point to be made:  the perturbation produced by the collision has a very characteristic geometry.  It has planar symmetry, is zero outside the surface of the lightcone of the collision, and at least near the surface increases linearly with distance inside.\footnote{We remind the reader again that this statement is valid only on angular scales larger than about a degree.  On smaller scales a more detailed analysis is required \cite{aniso}.}  The CMB temperature map provides a probe of a two dimensional slice of this three dimensional perturbation at decoupling.  As we have seen in this paper, $E$-mode polarization originates from and provides a probe of a {\em different} slice (or rather slices).  The information provided by polarization is independent---at least partially---of the information provided by temperature, and therefore the two together form a powerful check on the model.

Imagine that a temperature disk of the type predicted by \cite{wwc2} were to be observed in the CMB (and in fact a disk with at least roughly the correct profile does indeed appear to be present  \cite{coldspot1,coldspot2,coldspot3,coldspot4,coldspot5}).  Certainly such a feature could have been created by a random Gaussian perturbation---although perhaps with very low probability.  But with what probability would such a random fluctuation produce both the correct temperature map and the correct polarization signal to correspond to the one computed here?  The answer depends on many factors, among them uncertainties affecting the visibility function and reionization history of the Universe.  But it is clear that at least in principle, one can have considerable leverage with which to rule out a random origin for such a signal.

Another model that can produce cold or hot spots on the CMB temperature map are cosmic textures \cite{Turok:1990gw}.  Because textures are late decaying, they do not produce a significant polarization signal \cite{Cruz:2007pe}, and therefore can be distinguished from the collision model.  Similarly if the temperature perturbation is an integrated Sachs-Wolfe effect due to a void (or overdensity) at $z<5$ or so, its associated $E$-mode polarization signal will be minimal.

Of course the CMB is not the only tool available to study perturbations in the early Universe.  Other observations sensitive to the primordial power spectrum, such as the Lyman-alpha forest \cite{Weinberg:2003eg}, 21-cm radiation \cite{Loeb:2003ya,Kleban:2007jd}, and large-scale structure surveys (such as \cite{Abazajian:2008wr}) are also sensitive to the effects of bubble collisions \cite{wwc2, ktflow}.  Because of the special characteristics of the collision geometry, these probes should produce results that are correlated with the CMB in a specific way.  Since these measures access a three-dimensional volume, at least in principle far more data and far more statistical power remains to be tapped.  If indeed there is a feature in the CMB sky due to a bubble collision that is sufficiently bright to be observed, we are optimistic that a combination of these data would suffice to prove its nature beyond reasonable doubt.

We close with one final comment.  In this paper we have focused on the effects of a single collision.  But the results for the probability and distribution of characteristics of bubble collisions derived in \cite{bubmeas} had an interesting feature---the measure on the size of the collision disks is flat in $\mu_{c}=\cos \theta_{c}\propto x_{c}$.  The reason for this is fairly obvious in retrospect:  because of inflation, only a tiny region of the last scattering surface is visible to us today.  Collision lightcones slice through this surface at various positions.  While the distribution of their locations $x_{c,i}$ is not uniform, because we can only see a very small part of the surface, it is uniform to a very good approximation over the range we can observe.  Therefore disks with very small $\theta_{c}$ (and by the same token, disks with very small $\pi - \theta_{c}$) are unlikely.  The cold spot in the CMB has $\theta_{c}\sim10^{\circ}$ in angular radius.  Since the measure is flat in $\cos \theta_{c}$, the probability for this to be the only collision disk visible in our sky is small.  Instead, it may simply be the brightest....  which means that a search for larger, fainter disks is well worth doing \cite{disksearch}.

\section*{Acknowledgements}
We thank Olivier Dore, Gil Holder, Lam Hui, Adam Moss, Roman Scoccimarro, Stephen Shenker and Lenny Susskind for discussions.   The work of MK is supported by NSF CAREER grant PHY-0645435. The work of KS is supported in part by a NSERC of Canada Discovery grant.  The work of BC, KL and TSL is supported in part by Natural Sciences and Engineering Research Council of Canada and KL and TSL are also supported by the Institute of Particle Physics. KS thanks Perimeter Institute for Theoretical Physics, where part of this work was completed, for its hospitality.  TSL thanks the Center for Cosmology and Particle Physics at New York University for warm hospitality during part of the preparation of this work.  MK thanks the University of British Columbia for hospitality.

\bibstyle{apsrev4-1}

\bibliography{bubble}

\end{document}